\documentclass[runningheads]{llncs}

\usepackage[T1]{fontenc}
\def\doi#1{\href{https://doi.org/\detokenize{#1}}{\url{https://doi.org/\detokenize{#1}}}}
\usepackage{graphicx}
\usepackage{amsmath}
\usepackage{multirow}
\usepackage{booktabs}
\usepackage{soul,color}
\usepackage[table]{xcolor}
\usepackage{listings}
\usepackage{amssymb}
\begin{document}
\title{Multi-Feature Vision Transformer via Self-Supervised Representation Learning for Improvement of COVID-19 Diagnosis\thanks{Supported by organization x.}}
\author{Xiao Qi\inst{1} \and
	David J. Foran\inst{4} \and 
	John L. Nosher\inst{2} \and
	Ilker Hacihaliloglu	\orcidID{0000-0003-3232-8193}\inst{2,3}}

\institute{	Department of Electrical and Computer Engineering, Rutgers University, NJ, USA 
	\and 
	Department of Radiology, The University of British Columbia, BC, Canada
	\and
	Department of Medicine, The University of British Columbia, BC, Canada
	\and
	Rutgers Cancer Institute of New Jersey, NJ, USA}

\authorrunning{Qi et al.}

\maketitle              % typeset the header of the contribution
\begin{abstract}
	The role of chest X-ray (CXR) imaging, due to being more cost-effective, widely available, and having a faster acquisition time compared to CT, has evolved during the COVID-19 pandemic. To improve the diagnostic performance of CXR imaging a growing number of studies have investigated whether supervised deep learning methods can provide additional support. However, supervised methods rely on a large number of labeled radiology images, which is a time-consuming and complex procedure requiring expert clinician input. Due to the relative scarcity of COVID-19 patient data and the costly labeling process, self-supervised learning methods have gained momentum and has been proposed achieving comparable results to fully supervised learning approaches. In this work, we study the effectiveness of self-supervised learning in the context of diagnosing COVID-19 disease from CXR images. We propose a multi-feature Vision Transformer (ViT) guided architecture where we deploy a cross-attention mechanism to learn information from both original CXR images and corresponding enhanced local phase CXR images. We demonstrate the performance of the baseline self-supervised learning models can be further improved by leveraging the local phase-based enhanced CXR images. By using 10\% labeled CXR scans, the proposed model achieves 91.10\% and 96.21\% overall accuracy tested on total 35,483 CXR images of healthy (8,851), regular pneumonia (6,045), and COVID-19 (18,159) scans and shows significant improvement over state-of-the-art techniques. Code is available https://github.com/endiqq/Multi-Feature-ViT
	
\end{abstract}
\section{Introduction}
The rapid spread of COVID-19 outbreak caused a surge of patients to emergency departments and hospitalization. Compared to CT, chest X-ray (CXR) has several advantages such as its wide availability, exposure to less radiation, and faster image acquisition times. Due to this CXR has become the primary diagnostic tool for improved management of COVID-19. However, the interpretation of CXR images, compared to CT, is more challenging due to low image resolution and COVID-19 image features being similar to regular pneumonia. Computer-aided diagnosis via deep learning has been investigated to help mitigate these problems and help clinicians during the decision-making process \cite{qi2020chest,park2022multi,serena2021overview}. Most supervised deep learning methods rely on a large number of labeled radiology images. Medical image labeling is a time-consuming and complex procedure requiring expert clinician input.\\
\indent Semi-supervised learning methods have been proposed to provide a solution to the costly labeling process in the context of COVID-19 diagnosis. \cite{qi2021multi} proposed a multi-feature guided teacher-student distillation approach. Most recently self-supervised learning methods, which utilize all the unlabeled data during learning, have been investigated for COVID-19 diagnosis \cite{gazda2021self,hao2021self,park2021deep}. \cite{gazda2021self} achieved 79.5\% accuracy and 86.6\% area under the receiver operating characteristic curve (AUC) on 426 COVID-19 scans while using 1\% of the labeled data for the pretext task. \cite{hao2021self} reported a mean average precision of 41.6\% for 1,214 COVID-10 test data. \cite{park2021deep} reported 99.5\% accuracy using 607 COVID-19 scans. While these initial results on self-supervised learning are promising most of the prior work was evaluated on limited COVID-19 scans.\\
\indent In order to break the challenges associated with scarcity of training data and to boost classification performance, local phase-based CXR image enhancement has been proposed in \cite{qi2020chest,qi2021multi}. Motivated by this, in this work, we propose a new self-supervised learning approach where the proposed framework exploits local phase enhanced CXR image features to significantly improve the learning performance. Our contributions and findings include the following: 1) We developed MoCo-COVID, a Vision Transformer(ViT) with modified Momentum Contrast (MoCo) pretraining on CXR images for self-supervised learning. This is the first study pretraining a ViT using MoCo for COVID-19 diagnosis from CXR images. 2) We demonstrated the performance of MoCo-COVID can be significantly improved by leveraging the local phase-based enhanced CXR scans specially in low data regime. 80.27\% and 93.24\% overall accuracy were achieved tested on 799 and 14,123 COVID-19 scans while using 1\% of the labeled local phase-based enhanced data for the training. 3) A novel objective function was proposed using knowledge distillation to provide better generalization. 4) We developed a multi-feature ViT architecture based on cross-attention mechanism (MF-ViT CA) to further improve accuracy. The proposed MF-ViT CA achieves 95.03\% and 97.35\% mean accuracy on two large-sale test datasets including 14,922 COVID-19 scans and outperforms state-of-the-art semi-supervised learning and fully supervised learning methods.

%1) We propose a novel method for self-supervised learning, based on contrastive learning, where we leverage multiple images per subject to boost the model performance. Our method is a multi-feature Vision Transformer (ViT) guided architecture where we deploy a cross-attention mechanism to learn information from both original CXR images and corresponding enhanced local phase CXR images. 2) We pretrain a ViT with modified Momentum Contrast (MoCo)-pretraining on COVID-19 dataset and demonstrate MoCo-pretraining leads to a better performance than the model without a MoCo-pretraining for COVID-19 CXR interpretation. This is the first study pretraining a ViT using MoCo for COVID-19 diagnosis from CXR images. 3) We demonstrate the performance of the baseline self-supervised learning models can be further improved by leveraging the local phase-based enhanced CXR images. 4) We conduct a thorough experimental evaluation on 35,483 CXR images of healthy (8,851), regular pneumonia (6,045), and COVID-19 (18,159) scans achieving significant improvement over state-of-the-art techniques. To the best of our knowledge, this is the largest COVID-19 evaluation study reported for self-supervised learning.

\section{Methods}
\begin{figure}
	\centering                                  
	\includegraphics[width=12cm]{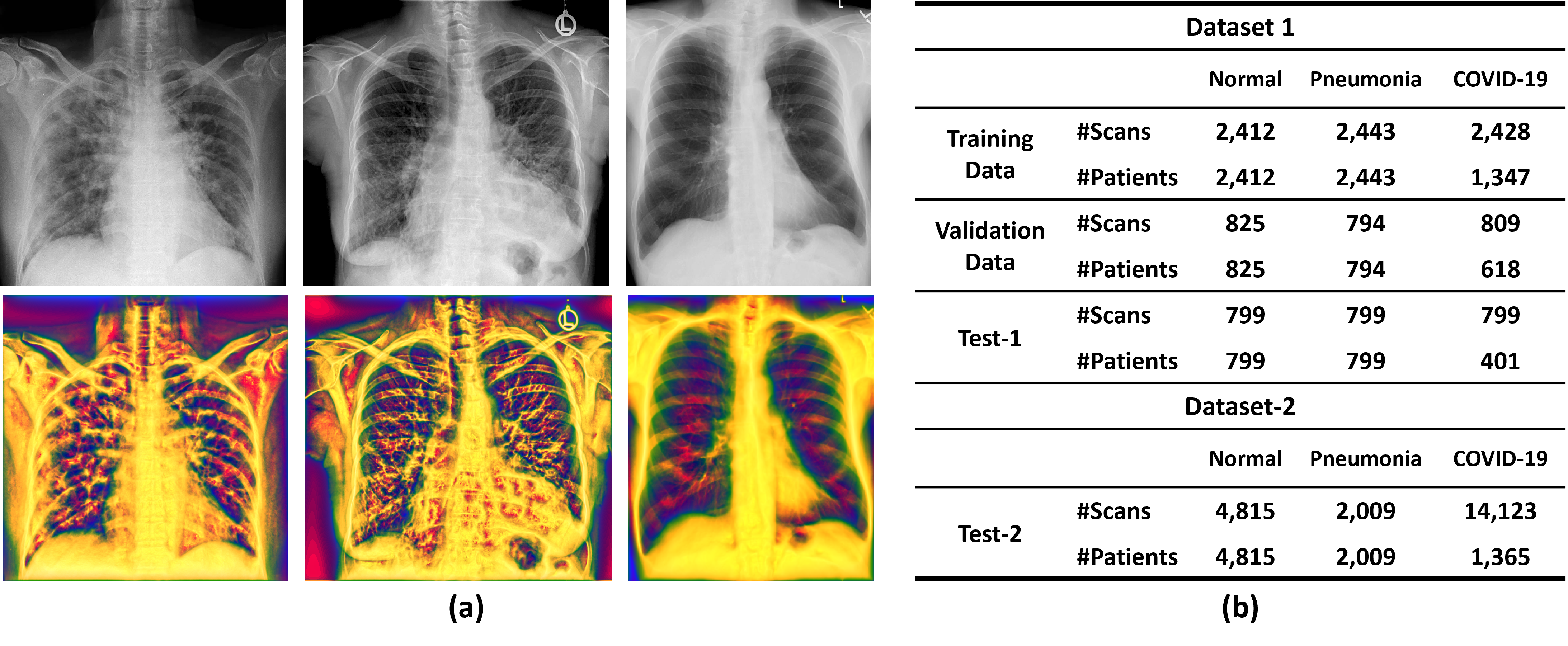}
	\caption{(a)Top row original CXR images. Bottom row $MF(x,y)$ images. The first two columns are from subjects who are diagnosed with COVID-19. The last column is from a healthy subject. (b) Class distribution of the evaluation datasets.}
	\label{fig:enhance} 
\end{figure}

\noindent\textbf{Datasets:} All images were collected from six public data repositories, which are BIMCV \cite{bimcv}, COVIDx \cite{Wang2020}, COVID-19-AR\cite{Rural}, MIDRC-RICORD-1c\cite{RSNA}, COVID-19 Image Repository\cite{Germany} and COVID-19-NY-SBU\cite{SBU}. Our dataset consists of a total of 33,055 CXR scans from 18,252 patients with three classes: normal, pneumonia, and COVID-19. These images were split into two datasets: \textbf{1) Dataset-1:} A subset containing 12,108 CXR scans with a balanced distribution of classes (Fig. \ref{fig:enhance}(b)). Dataset-1 was split into 60\% training, 20\% as validation, and 20\% as testing dataset. No subject overlaps among train, validation, and test datasets. The test data in this dataset is referred to as \textit{Test-1}. \textbf{2) Dataset-2:} Includes 20,947 CXR scans from 8,181 patients and has a larger number of scans in COVID-19 class. Dataset-2 only serves as an additional test dataset, referred to as \textit{Test-2}, for evaluating the robustness of the proposed methods. 

\noindent\textbf{Image Enhancement:} Local phase-based image analysis methods are more robust to intensity variations, usually arising from patient characteristics or image acquisition settings, and have been incorporated into various medical image processing tasks\cite{alessandrini2012myocardial,li2018hybrid,zhao2015retinal}. The enhanced local phase CXR image, denoted as $MF(x,y)$, is obtained by combining three different local phase image features: 1-Local weighted mean phase angle ($LwPA(x,y)$), 2- Weighted local phase energy ($LPE(x,y)$), and 3- Enhanced local energy attenuation image ($ELEA(x,y)$). $LPE(x,y)$ and $LwPA(x,y)$ image features are extracted by filtering the CXR image in frequency domain using monogenic filter and $\alpha$-scale space derivative (ASSD) bandpass quadrature filters \cite{qi2020chest}. $ELEA(x,y)$ image is extracted, processing the $LPE(x,y)$ image, by modeling the scattering and attenuation effects of lung tissue inside a local region using L1 norm-based contextual regularization method \cite{qi2020chest}. We have used the filter parameters reported in \cite{qi2020chest} for enhancing all the CXR images. Investigating Figure. \ref{fig:enhance}(a) we can see that structural features inside the lung tissue are more dominant for COVID-19 CXR images compared to healthy lung (last column Fig.\ref{fig:enhance}(a)). The enhanced local phase CXR images ($MF(x,y)$) and the original CXR images are used to train proposed self-supervised learning methods explained in the next sections.

\begin{figure}
	\begin{center}
		\includegraphics[width=12cm]{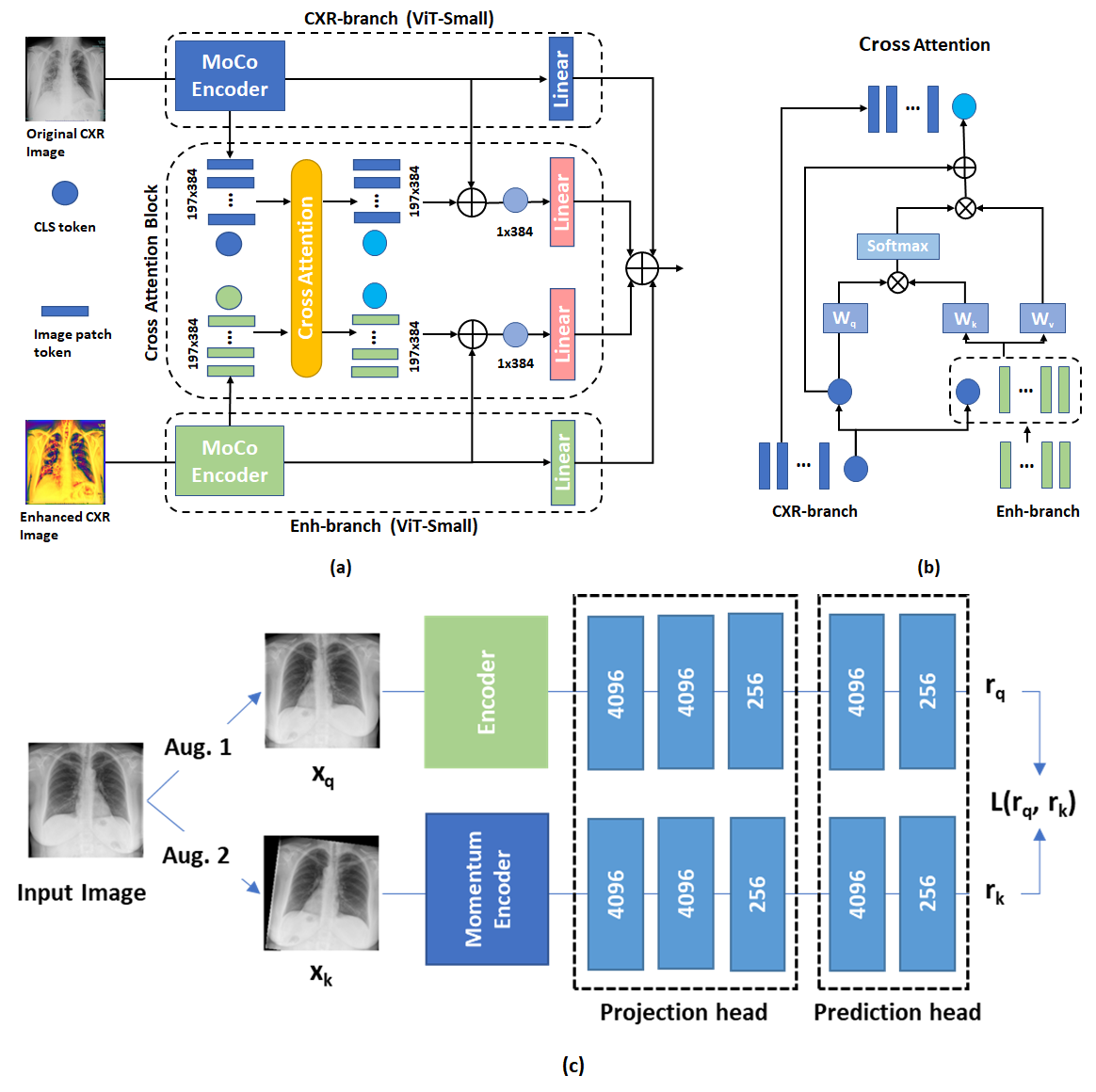}
		\caption{ (a) Proposed Multi-Feature Fusion ViT architecture using the cross attention block which is shown in (b). (c) The proposed MoCo-COVID architecture.}
		\label{pipeline}
	\end{center}
\end{figure}

\noindent\textbf{MoCo-COVID-Self-Supervised Pretraining using MoCo:}\label{MoCo} We introduce a MoCo-COVID framework (Fig.\ref{pipeline}(c)). During the self-supervised training, a CXR image is first transformed via two random augmentations(Aug.1 and Aug.2) into images $ x_q$ and $x_k$. $ x_q $ is passed through an encoder network, while $ x_k $ is passed through a momentum encoder network. We choose ViT-Small (ViT-S) \cite{vit} with 6 heads as the backbone, instead of 12 heads used by MoCo v3, to have a lower parameter count, and a faster throughput \cite{mocov3}. Sine-cosine variant \cite{attention} is added to the sequence as positional embedding. The representations generated by each network are then passed into the projection head followed by a prediction head. The projection head has three layers and the prediction head has two layers. Each layer follows with batch normalization (BN) and a rectified linear unit (ReLU) except the last layer of both projection head and prediction head. Then the InfoNCE contrastive loss function \cite{van2018representation} is adopted to promote the similarity between the representations $r_q$ and $r_k$:
\begin{equation} \label{eq:1_}
	L(r_q, r_k) = -log\dfrac{\mathrm{exp}(r_q.r_k+/\tau)}{\sum_{i=1}^{K}\mathrm{exp}(r_q.r_k,i/\tau)}
\end{equation} 
, where $ \tau $  is a temperature hyperparameter and K is the number of currently stored representations. The momentum coefficient follows a cosine schedule changing from 0.9 to 0.999 during the MoCo-COVID pretraining.

\noindent\textbf{Multi-Feature Fusion Vision Transformer via Cross Attention:} Figure.\ref{pipeline}-(a) illustrates the architecture of our proposed Multi-Feature Vision Transformer with cross-attention (CA) block (MF-ViT CA), which consists of two branches and a CA block. \textit{CXR-branch} is used for processing the original CXR image, \textit{Enh-branch} is used for processing the local phase enhanced CXR image ($MF(x,y)$), and CA block for extracting information from both branches. During the forward pass, an original CXR image and corresponding $MF(x,y)$ image is first passed to the pretrained MoCo encoder in parallel to obtain a tensor of dimension 197 $\times$ 384 for each image. These two tensors are then fed as inputs to our CA block. In the CA block, the CLS token of one branch fuses with patch tokens of the other branch using CA mechanism, and after fusion the CLS token concatenates with its own patch tokens again to produce an output in the same dimension of input tensor for each branch, described more details in below. The outputs of cross-attention and MoCo encoder are then combined via element-wise summation for each branch. The CLS token of dimension 1 $\times$ 384 from each branch as a compact representation, which encodes the information from both the original CXR image and $MF(x,y)$ image, is fed into a linear projection layer with three units. After that, the outputs of the two projection functions are fused using an element-wise summation.

\noindent\textbf{Loss Function:} 
We used hard-label distillation, which is a variation of distillation introduced by \cite{deit}, for training CA block. In our study, \textit{CXR-branch} and \textit{Enh-branch}, which are obtained with the MoCo-COVID-pretrained ViT-S by using original and enhanced CXR scans(denoted as $ cxr $ and $ enh $ in Eq.\ref{eq:1}), are considered as the teacher models. The CA block is considered the student model. The objective function with this hard-label distillation is:

\begin{equation} \label{eq:1}
	L^{\mathrm{hardDistill}}_{\mathrm{global}}   = \frac{1}{3}L_{\mathrm{CE}}(\psi(Z_{s}),y) + \frac{1}{3}L_{\mathrm{CE}}(\psi(Z_{s}),y_{t_{cxr}})+\frac{1}{3}L_{\mathrm{CE}}(\psi(Z_{s}),y_{t_{enh}}).
\end{equation}
where $ Z_s $ is the logits of the student model and $ \psi $ is the softmax function. The idea is to use both the real target $ y $ and the target generated by the teachers $y_t=\mathrm{argmax}_cZ_t(c)$, where $ Z_t $ is the logits of the teacher models. For a given image, the predicated label associated with the teacher may change due to data augmentation during the forward pass. Thus, the teacher model is aiming at producing predicted labels that are similar but not identical to true label.
% , denoted as $ enh $
%\in \mathbb{R}^{N \times D} \), where N is the number of tokens in the sequence and each token is represented by a feature vector $ D $,

The major benefit of the proposed hard label distillation is being parameter-free compared to soft distillation in \cite{soft_dist}. In our study, the probability distribution has the correct class at a very high probability, with all other class probabilities close to 0. The problem with these weak probabilities is they do not capture desirable information for the student model to learn effectively. To address this issue, we proposed this hard label distillation loss function by adding two distillation losses, which are losses between the predicted label of teacher model and logits of student model. Distillation loss would soften the distributions predicted by the teacher model so that the student model can learn more information, and this is especially useful when dealing with small datasets \cite{soft_dist}. And the proposed objective function ensures the student model inherits better quality from the teacher model and mitigates the over-confidence issue of neural networks by improving the generalization.

\noindent\textbf{Cross-Attention (CA) Mechanism:} The proposed CA mechanism is inspired by \cite{chen2021crossvit}, which is designed for fusing multi-scale features from two branches. In our study, we intend to extract information from two types of images via cross-attention, and thus we removed projection function used in \cite{chen2021crossvit} for changing the dimension of feature maps. Figure-\ref{pipeline}(b) presents the structure of our proposed cross-attention. Similar to \cite{chen2021crossvit}, we utilize the CLS token at each branch to fuse with patch tokens from the other branch and then project it back to its own branch in order to exchange information among the patch tokens from the other branch. Since the CLS token already learns the extracted information from all patch tokens in its own branch, interacting with the patch tokens from the other branch helps to include information from different feature inputs. In the following section, we provide a detailed explanation about the CA mechanism for the \textit{CXR-branch}. \textit{Enh-branch} follows the same process.

\noindent\textbf{CXR-branch-CA:} An illustration of CA for \textit{CXR-branch} is presented in the Figure \ref{pipeline}(b). Firstly, the CLS tokens from \textit{CXR-branch}, denoted as $ cxr $, concatenates the patch tokens from the \textit{Enh-branch} to form \(x^{'cxr}\) shown as Eq.\ref{eq:1-}.
\begin{equation} \label{eq:1-}
	x ^{'cxr}   = [\ x^{cxr}_{cls}\ \|\ x^{enh}_{patch}\ ]\,
\end{equation} 
Then, the CA is performed between $x^{cxr}_{cls}$ and $x ^{'cxr} $ using linear projections to computer queries, keys and values (\textbf{Q}, \textbf{K} and \textbf{V}), where CLS token is the only information used in query. And it uses the scaled dot product for calculating the attention weights between $ \textbf{Q} $ and $\textbf{ K} $ and then aggregates $ \textbf{V} $. The CA can be expressed as below:

\begin{gather} 
	\textbf{Q} = x^{cxr}_{cls}\textbf{W}_q, \ \textbf{K} = x^{'cxr}\textbf{W}_k, \ \textbf{V}=x^{'cxr}\textbf{W}_v,  \nonumber \\ 
	\textbf{CA} = \mathrm{softmax}(\textbf{QK}^T/ \sqrt{D/h})\textbf{V}, 
	\label{eq:2-}
	%\ CA(x^{'ori}) = %\textbf{AV}
\end{gather}

In the Eq.\ref{eq:2-}, $\textbf{W}_q $,$ \textbf{ W}_k $,$  \textbf{W}_v $ \(\in \mathbb{R}^{C \times (C/h)} \) are learnable parameters, where C and h are the embedding dimension and number of heads. We use the three heads in the cross attention in this study. In the end, the new CLS token of \textit{CXR-branch}, which is obtained by cross attention and residual shortcut, concatenates with patch tokens from \textit{CXR-branch} as the output of \textit{CXR-branch} shown as below:
\begin{equation} \label{eq:3}
	y^{cxr}_{cls} = x^{cxr}_{cls} + \textbf{CA}, \
	z^{cxr} = [\ y^{cxr}_{cls}\ \| \ x^{cxr}_{patch}\ ]
\end{equation}
%We denote the input sequence from original CXR image as \(\textbf{F}^{in} \in \mathbb{R}^{N \times D_f} \). Then the cross-attention (CA) was performed  
%
%\(x ^{'ori} = [ x^{ori}_{cls} || x^{enh}_{patch} ]\)

%, where the CLS token from one branch are fused with class tokens from the other branch. 
%of transformer \cite{attention}  as in self-attention
%We pretrained models on different fractions of training data. Label fraction represents the ratio of labeled data retained for finetuning. For  
%We applied both supervised and self-supervised pretraining procedures. We evaluated Vit-Small \cite{vit} and ResNet50 \cite{res} for supervised learning. We closely followed the training procedures reported in \cite{qiijcar} for pretraining ResNet50\cite{res}. To be specific, we use SGD optimizer with a learning rate of 10$ ^-3 $ and batch size of 16. Vit-Small model was pretrained with SGD optimizer with a learning rate of 6$^-4$ and batch size of 16. and \cite{mocov3} for pretraining ResNet50\cite{res} to keep the same rate reported in \cite{mocov3} 
\section{Experiments and Results}
\noindent\textbf{MoCo-COVID Pretraining:} We pretrained our MoCo-COVID end-to-end for both original CXR image (denoted as CXR-ViT-S) and $MF(x,y)$ (denoted as Enh-ViT-S), on all training dataset without label information. 
MoCo-COVID pretraining initialization was performed using the weights obtained on ImageNet-initialized models for a faster convergence \cite{transfer}. Data augmentation included resizing to a 224 $\times$ 224 gird, random rotation (10 degrees), and horizontal flipping similar to \cite{mococxr}. We maintained hyperparameters related to momentum, weight decay, and feature dimension from MoCo v3 \cite{mocov3}. To be specific, the model was optimized by AdamW \cite{adamw}, a weight decay parameter of 0.1 and batch size of 16 on 2 NVIDIA GTX 1080 using PyTorch's DistributedDataParallel framework \cite{parallel}. We pretrained for 300 epochs (40 warm-up epochs) with a cosine linear-rate scheduler and set the initial learning rate as $ lr \times$ BatchSize$ /4 $ , where $ lr $ is $ 1.5e^{-4} $. \\
\indent We fine-tuned models with different fractions of labeled data. Label fraction represents the percentage of labeled data retained during fine-tuning. For example, a model fine-tuned with 1$ \% $ label fraction meaning the model will only have access to 1$ \% $ of the training dataset as labeled dataset, and the remaining 99$ \% $ are hidden from the model as unseen data. The label fractions of training dataset are 0.25$ \% $(18 scans), 1$ \% $ (72 scans), 10$ \% $(728 scans), 30$ \% $(2184 scans), and 100$ \% $. Fine-tuning was repeated five times for each label fraction. Label fractions less than 100$ \% $ are random samples from the training dataset.\\
\indent We conducted two fine-tuning ablations, which are linear probing (LP) and end-to-end fine-tuning (FT). LP means the pretrained weight values of the MoCo-COVID encoder were frozen and, after removing the projection and prediction heads, a new linear classifier with randomly reinitialized weights was added and fine-tuned using labeled data. FT allowed the entire model including MoCo-COVID encoder to fine-tune not just the newly added classifier. The models were fine-tuned using 90 epochs. All fine-tunings used the cosine annealing learning rate decay\cite{cosannealing} and SGD \cite{adam} optimizer.

\noindent\textbf{Ablations of Multi-Feature ViT using cross-attention mechanism (MF-ViT CA):} The weights of the \textit{CXR-branch} and \textit{Enh-branch} are initialized to the MoCo-COVID pretrained weight values of CXR-ViT-S FT and Enh-ViT-S FT respectively, and the weights of the CA block are randomly reinitialized with a uniform distribution \cite{normaldistribution}. The CA block was fine-tuned using labeled data with hard-label distillation. We compared the proposed MF-ViT CA with a model, where it also has two MoCo-COVID pretrained branches (\textit{CXR branch} and \textit{Enh branch}) without using cross-attention block, denoted as MF-ViT LP. During the fine-tuning, we only fine-tuned the linear layers of \textit{CXR-branch} and \textit{Enh-branch}. \\
\noindent\textbf{Baselines:} As baseline comparison we report results from MoCo-CXR \cite{mococxr}, which uses MoCo v2 pretrained Dense121, when trained using original CXR and local phase enhanced CXR images ($MF(x,y)$). We report end-to-end fine-tuning protocol (FT) results for MoCo-CXR \cite{mococxr} and the network architecture was optimized to achieve the best results using our dataset to provide a fair comparison. Finally, we also compare our results against \cite{qi2021multi} semi-supervised, and fully supervised methods where $MF(x,y)$  were used. \\
\noindent\textbf{Quantitative Results:} Quantitative results are displayed in Table \ref{tab:sl_ssl_1}. The proposed MF-ViT CA architecture achieved the best accuracy compared to the rest of the self-supervised learning models when label fractions were more than 1\%. MF-ViT CA performed significantly better (paired t-test p$<$0.05) when tested on \textit{Test-2} data compared to \textit{Test-1} data proving the robustness of the method when tested on large scale COVID-19 data (14,123 images). The proposed Enh-ViT-S FT model had the highest accuracy for label fraction less than 10\%. We can also observe that end-to-end fine-tuning protocol (FT) results are significant improvements over linear probing (LP) protocol in the proposed MoCo-COVID model. The accuracy of the baseline architecture MoCo-CXR \cite{mococxr} and all the proposed architectures improve when enhanced local phase CXR images ($MF(x,y)$) were used as an input. Finally, we have also observed that the CA mechanism significantly improves the results of the proposed MF-ViT LP model(paired t-text p$<$0.05) (Table \ref{tab:sl_ssl_1}). From the results, we observe that the MF-ViT CA yields a statistically significant gain (paired t-text p$<$0.05) compared to MF-TS \cite{qi2021multi} at 0.25\% (18 scans) and 1\% label fractions (72 scans). This indicates that the proposed models provide high-quality representations, better generalization capability, and transferable initialization for COVID-19 interpretation for minimal label fractions and when evaluated on large test data.

%However, the \textit{Test-1} and \textit{Test-2} data used in \cite{qi2021multi} was 6\% and 52\% less respectively compared to the current work. Additional quantitative evaluation results can be found in the supplementary file. 

%The MF-ViT CA outperforms the rest all models when label fractions are more 1\% shown as Fig.\ref{results}. In the Test-1 of Fig.\ref{results}(b), the MF-ViT CA with 10\% label fraction achieved an accuracy of 91.10\% while the second best Dense121\cite{dense} FT achieved an Acc of 90.23\%, yielding a statistically significant Acc improvement of 0.87\% ($ p $<0.05 with paired t-test). In the Test-2 of Fig.\ref{results}(b), the MF-ViT CA model performed better than the Enh-ViT-S FT, which has the second best accuracy across all models, at 10\%, 30\%, and 100\% label fraction and gains statistical significance at 30\% and 100\% label fraction ($ p $<0.05 with paired t-test). The Enh-ViT-S FT model had the highest accuracy with label fraction less than 10\%. Additional quantitative evaluation results can be found in the supplementary file.

\begin{table}[]
	\begin{center}
		\caption{Accuracy results obtained from \textit{Test-1} and \textit{Test-2} data. Green shaded region corresponds to the highest scores obtained. * indicates statistical improvement compared with second best self-supervised learning method (p$<$0.05) using paired t-test. }		
		\label{tab:sl_ssl_1}
			\centering \scalebox{0.85}{
		\begin{tabular}{c cc cc cc cc cc  }
		\toprule
		%			\hline 
		\multirow{2}{*}{\textbf{Method}}  & \multicolumn{2}{c}{\textbf{ 0.25\%}}  & \multicolumn{2}{c}{\textbf{ 1\% }} & 
		\multicolumn{2}{c}{\textbf{ 10\% }} & \multicolumn{2}{c}{\textbf{ 30\% }} &\multicolumn{2}{c}{\textbf{ 100\%}}\\
		
		\cmidrule(lr){2-3}\cmidrule(lr){4-5}\cmidrule(lr){6-7}\cmidrule(lr){8-9}\cmidrule(l){10-11}
		
		&  Test1 & Test2 & Test1  & Test2 & Test1 & Test2 & Test1 & Test2& Test1 & Test2\\
		
		%			\midrule
		%			&\multicolumn{10}{c}{\textbf{ImageNet Pretrained ViT}}\\
		%			\midrule
		%			\multicolumn{1}{l}{ViT-S\cite{vit} LP} 
		%			& 55.72 & 55.34 & 73.69 & 68.51 & 82.25 & 80.18 & 84.42  & 82.68  & 85.72  & 85.72\\
		%			%			& 55.34 & 0.756 & 68.51 & 0.906 & 80.35 & 0.953 & 82.68  & 0.955  & 85.72  & 0.970\\		
		%			\multicolumn{1}{l}{ViT-S\cite{vit} FT}
		%			& 59.23 & 61.88 & 76.37 & 69.75 & 88.10 & 82.19 & 91.61 & 89.08 & 92.55 & 91.79\\
		%			& 61.88 & 0.806 & 69.75 & 0.918 & 84.57 & 0.971 & 89.08 & 0.982 & 91.79 & 0.990\\
		\midrule
		&\multicolumn{10}{c}{\textbf{ MoCo-CXR\cite{mococxr}-pretrained end-to-end Dense121}}\\
		\midrule
		\multicolumn{1}{l}{FT CXR}    
		& 65.16 & 73.85 & 78.85 & 77.52 & 90.23 & 84.57 & 92.86 & 90.62 & 94.74 & 93.70\\
		\multicolumn{1}{l}{FT Enh}    
		& 73.60 & 81.85 & 83.12 & 90.87 & 90.40 & 95.84 & 91.57 & 95.95 & 93.73 & 96.50\\
		%			& 73.85&0.892 & 77.52& 0.935& 82.19& 0.962& 90.62& 0.984& 93.70& 0.991\\									                       
		
		\midrule
		&\multicolumn{10}{c}{\textbf{ MoCo-COVID Pretrained (ours)}}\\	 			
		\midrule
		\multicolumn{1}{l}{CXR-ViT-S LP}     
		& 72.64 & 74.18 & 77.35 & 78.33 & 83.76 & 81.71 & 85.10 & 83.88 & 86.93 & 86.15\\
		\multicolumn{1}{l}{Enh-ViT-S LP}     
		& 79.97 & 91.71 & 84.06 & 94.00 & 87.81 & 94.87 & 89.10 & 95.85 & 90.51 & 94.14\\
		\multicolumn{1}{l}{CXR-ViT-S FT}    
		& 73.00 & 73.37 & 78.23 & 81.13 & 88.59 & 85.94 & 91.66 & 89.48 & 93.26 & 92.19\\
		%\hline
		\multicolumn{1}{l}{Enh-ViT-S FT}    
		&\cellcolor{green!25}\textbf{80.27 }& \cellcolor{green!25}\textbf{93.24*} & \cellcolor{green!25}\textbf{84.10} & \cellcolor{green!25}\textbf{94.00} & 89.09 & 95.33 & 91.43 & 96.23 & 92.62 & 96.57\\
		%			&93.24 & 0.976 & 93.98 & 0.987 & 95.33 & 0.993 & 96.23 & 0.994 & 96.57 & 0.994\\
		\midrule
		&\multicolumn{10}{c}{\textbf{ Multi-Feature Model (ours)}}\\	
		\midrule
		\multicolumn{1}{l}{MF-ViT LP}     
		& 69.01 & 65.22 & 67.20& 69.72& 86.72& 85.20& 91.06& 92.02& 92.53& 95.01\\
		%			& 65.22 & 0.854 & 69.72& 0.899& 85.20& 0.968& 92.02& 0.985& 95.01& 0.990\\
		%\hline
		\multicolumn{1}{l}{MF-ViT CA}  
		& 79.88 & 89.91 & 82.72 & 92.57 & \cellcolor{green!25}\textbf{91.10*} & \cellcolor{green!25}\textbf{96.21 }& \cellcolor{green!25}\textbf{93.27} & \cellcolor{green!25}\textbf{96.84*} & \cellcolor{green!25}\textbf{95.03} & \cellcolor{green!25}\textbf{97.35*}\\
		
		\midrule
		&\multicolumn{10}{c}{\textbf{Semi-Supervised Learning}}\\	 			
		\midrule
		\multicolumn{1}{l}{MF-TS\cite{qi2021multi}}    
		& 77.13 & 80.93 & 82.27 & 86.57 & 90.73 & 95.65 & 92.68 & 96.35 & - & - \\
		\midrule
		&\multicolumn{10}{c}{\textbf{Fully-Supervised Learning}}\\	 			
		\midrule
		\multicolumn{1}{l}{XNet\cite{chollet2017xception}}    
		& - & - & - & - & - & - & - & - & 94.38 & 89.20\\
		\multicolumn{1}{l}{InceptionV4\cite{szegedy2016inceptionv4}}    
		& - & - & - &  -& - & - &  -&  -& 93.98 & 88.92\\
		
		\bottomrule
		\end{tabular}}
	\end{center}
\end{table}

\section{Conclusion}
Our large quantitative evaluation results obtained using the largest COVID-19 data collected from different sites, show the significant improvements achieved using the local phase image features for self-supervised learning. Although we did not have access to the CXR machine type and non-image patient information (BMI, age, sex) we believe the large data used in this work represents images with varying image acquisition settings and intensity variations. Our quantitative results show significantly improved accuracy values over the investigated baselines proving the robustness of our proposed methods. Future work will include the extension of the method for diagnosing different lung diseases from CXR images. 

\section{Appendix}

\subsection{Quantitative evaluation: Additional results and comparison against fully-supervised learning and semi-supervised learning}
Precision, recall and F1-scores for \textit{Test-1} and \textit{Test-2} data for all the investigated methods are shown in Table \ref{tab:sl_ssl_2}. Table \ref{tab:sl_ssl_1} displays accuracy results for all the investigated methods. Investigating the results it can be observed that the proposed Enh-ViT-S FT self-supervised learning method achieves best performance when the label fraction of training dataset are \%0.25 (18 scans) and \%1 (72 scans). When trained in \%100 label fraction our MF-ViT CA method also outperforms fully supervised baselines (Tables \ref{tab:sl_ssl_1}-\ref{tab:sl_ssl_2}).
\begin{table}[]
	\begin{center}
		\caption{Recall and precision results obtained from \textit{Test-1} and \textit{Test-2} data. Green shaded region corresponds to the highest scores obtained.}		
		\label{tab:sl_ssl_2}
		\begin{tabular}{c c cc cc cc cc cc  }
			\toprule
			%			\hline 
			\multirow{2}{*}{\textbf{Method}} & \multirow{2}{*}{\textbf{Metrics}} & \multicolumn{2}{c}{\textbf{ 0.25\%}}  & \multicolumn{2}{c}{\textbf{ 1\% }} & 
			\multicolumn{2}{c}{\textbf{ 10\% }} & \multicolumn{2}{c}{\textbf{ 30\% }} &\multicolumn{2}{c}{\textbf{ 100\%}}\\
			
			\cmidrule(lr){3-4}\cmidrule(lr){5-6}\cmidrule(lr){7-8}\cmidrule(lr){9-10}\cmidrule(l){11-12}
			
			& &  Test1 & Test2 & Test1  & Test2 & Test1 & Test2 & Test1 & Test2 & Test1 & Test2\\
			
			\midrule
			&&\multicolumn{10}{c}{\textbf{ImageNet Pretrained ViT}}\\
			\midrule
			\multicolumn{1}{l}{\multirow{3}{*}{ViT-S\cite{vit} LP}}
			
			&\multicolumn{1}{l}{Precision} & 0.58 & 0.71 & 0.75 & 0.86 & 0.83 & 0.90 & 0.85  & 0.89  & 0.86  & 0.91\\
			&\multicolumn{1}{l}{Recall}    & 0.56 & 0.55 & 0.74 & 0.69 & 0.82 & 0.80 & 0.84  & 0.83  & 0.86  & 0.86\\
			&\multicolumn{1}{l}{F1-Score}  & 0.55 & 0.60 & 0.74 & 0.74 & 0.82 & 0.83 & 0.84  & 0.85  & 0.86  & 0.87\\
			
			\multicolumn{1}{l}{\multirow{3}{*}{ViT-S\cite{vit} FT}}
			&\multicolumn{1}{l}{Precision} & 0.61 & 0.76 & 0.77 & 0.87 & 0.88 & 0.91 & 0.92 & 0.93 & 0.93 & 0.94\\
			&\multicolumn{1}{l}{Recall}    & 0.59 & 0.62 & 0.76 & 0.70 & 0.88 & 0.82 & 0.92 & 0.89 & 0.93 & 0.92\\
			&\multicolumn{1}{l}{F1-Score}  & 0.58 & 0.66 & 0.76 & 0.75 & 0.88 & 0.85 & 0.92 & 0.90 & 0.93 & 0.93\\
			
			\midrule
			& &\multicolumn{10}{c}{\textbf{ MoCo-CXR\cite{mococxr}-end-to-end pretrained DenseNet121}}\\
			\midrule			
			\multicolumn{1}{l}{\multirow{3}{*}{FT CXR}}    
			& \multicolumn{1}{l}{Precision} & 0.66 & 0.81 & 0.79 & 0.88 & 0.90 & 0.92 & 0.93 & 0.93 & 0.95 & 0.95\\
			& \multicolumn{1}{l}{Recall}    & 0.65 & 0.74 & 0.79 & 0.78 & 0.90 & 0.85 & 0.93 & 0.91 & 0.95 & 0.94\\
			& \multicolumn{1}{l}{F1-Score}  & 0.65 & 0.77 & 0.79 & 0.81 & 0.90 & 0.87 & 0.93 & 0.91 & 0.95 & 0.94\\
			
			\multicolumn{1}{l}{\multirow{3}{*}{FT Enh}}    
			& \multicolumn{1}{l}{Precision} & 0.74 & 0.86 & 0.84 & 0.94 & 0.91 & 0.96 & 0.92 & 0.96 & 0.94 & 0.97\\
			& \multicolumn{1}{l}{Recall}    & 0.74 & 0.82 & 0.83 & 0.91 & 0.90 & 0.96 & 0.92 & 0.96 & 0.94 & 0.97\\
			& \multicolumn{1}{l}{F1-Score}  & 0.74 & 0.84 & 0.83 & 0.92 & 0.90 & 0.96 & 0.92 & 0.96 & 0.94 & 0.97\\
			
			\midrule
			& &\multicolumn{10}{c}{\textbf{ MoCo-COVID Pretrained - Ours}}\\	 			
			\midrule
			\multicolumn{1}{l}{\multirow{3}{*}{CXR-ViT-S LP}}     
			& \multicolumn{1}{l}{Precision} & 0.73 & 0.84 & 0.77 & 0.88 & 0.84 & 0.89 & 0.85 & 0.90 & 0.87 & 0.90\\
			& \multicolumn{1}{l}{Recall}    & 0.73 & 0.74 & 0.77 & 0.78 & 0.84 & 0.82 & 0.85 & 0.84 & 0.87 & 0.86\\
			& \multicolumn{1}{l}{F1-Score}  & 0.73 & 0.77 & 0.77 & 0.81 & 0.84 & 0.84 & 0.85 & 0.86 & 0.87 & 0.87\\
			\multicolumn{1}{l}{\multirow{3}{*}{CXR-ViT-S FT}}    
			& \multicolumn{1}{l}{Precision} & 0.81 & 0.92 & 0.85 & 0.95 & 0.88 & 0.95 & 0.89 & 0.96 & 0.90 & 0.94\\
			& \multicolumn{1}{l}{Recall}    & 0.80 & 0.92 & 0.84 & 0.94 & 0.88 & 0.95 & 0.89 & 0.96 & 0.90 & 0.91\\
			& \multicolumn{1}{l}{F1-Score}  & 0.80 & 0.92 & 0.84 & 0.94 & 0.88 & 0.95 & 0.89 & 0.96 & 0.90 & 0.92\\
			
			\multicolumn{1}{l}{\multirow{3}{*}{Enh-ViT-S LP}}     
			&\multicolumn{1}{l}{Precision}  & 0.84 & 0.84 & 0.88 & 0.88 & 0.89 & 0.92 & 0.92 & 0.93 & 0.93 & 0.94\\
			&\multicolumn{1}{l}{Recall}     & 0.73 & 0.73 & 0.78 & 0.81 & 0.89 & 0.86 & 0.92 & 0.89 & 0.93 & 0.94\\
			&\multicolumn{1}{l}{F1-Score}   & 0.77 & 0.77 & 0.81 & 0.83 & 0.89 & 0.88 & 0.92 & 0.91 & 0.93 & 0.94\\
			
			\multicolumn{1}{l}{\multirow{3}{*}{Enh-ViT-S FT}}    
			& \multicolumn{1}{l}{Precision} & \cellcolor{green!25}\textbf{0.81} & \cellcolor{green!25}\textbf{0.93} & \cellcolor{green!25}\textbf{0.85} & \cellcolor{green!25}\textbf{0.95} & 0.89 & 0.96 & 0.92 & 0.97 & 0.93 & 0.97\\
			& \multicolumn{1}{l}{Recall}    & \cellcolor{green!25}\textbf{0.80} & \cellcolor{green!25}\textbf{0.93} & \cellcolor{green!25}\textbf{0.84} & \cellcolor{green!25}\textbf{0.94} & 0.89 & 0.95 & 0.91 & 0.96 & 0.93 & 0.97\\
			& \multicolumn{1}{l}{F1-Score}  & \cellcolor{green!25}\textbf{0.80} & \cellcolor{green!25}\textbf{0.93} & \cellcolor{green!25}\textbf{0.84} & \cellcolor{green!25}\textbf{0.94} & 0.89 & 0.95 & 0.91 & 0.96 & 0.93 & 0.97\\
			
			\midrule
			& &\multicolumn{10}{c}{\textbf{ Multi-Feature Model - Ours}}\\	
			\midrule
			\multicolumn{1}{l}{\multirow{3}{*}{MF-ViT LP}}     
			&\multicolumn{1}{l}{Precision} & 0.70 & 0.78 & 0.73 & 0.87 & 0.88 & 0.93 & 0.91 & 0.94 & 0.93 & 0.96\\
			&\multicolumn{1}{l}{Recall}    & 0.69 & 0.65 & 0.67 & 0.70 & 0.87 & 0.85 & 0.91 & 0.92 & 0.93 & 0.95\\
			&\multicolumn{1}{l}{F1-Score}  & 0.69 & 0.70 & 0.66 & 0.75 & 0.87 & 0.87 & 0.91 & 0.93 & 0.93 & 0.95\\
			
			\multicolumn{1}{l}{\multirow{3}{*}{MF-ViT CA}}  
			& \multicolumn{1}{l}{Precision} & 0.80 & 0.91 & 0.83 & 0.94 & \cellcolor{green!25}\textbf{0.91} & \cellcolor{green!25}\textbf{0.97} & \cellcolor{green!25}\textbf{0.93} & \cellcolor{green!25}\textbf{0.97} & \cellcolor{green!25}\textbf{0.95} & \cellcolor{green!25}\textbf{0.98}\\
			& \multicolumn{1}{l}{Recall}    & 0.80 & 0.90 & 0.83 & 0.93 & \cellcolor{green!25}\textbf{0.91} & \cellcolor{green!25}\textbf{0.96} & \cellcolor{green!25}\textbf{0.93} & \cellcolor{green!25}\textbf{0.97} & \cellcolor{green!25}\textbf{0.95} & \cellcolor{green!25}\textbf{0.97}\\
			& \multicolumn{1}{l}{F1-Score}  & 0.80 & 0.90 & 0.83 & 0.93 & \cellcolor{green!25}\textbf{0.91} & \cellcolor{green!25}\textbf{0.96} & \cellcolor{green!25}\textbf{0.93} & \cellcolor{green!25}\textbf{0.97} & \cellcolor{green!25}\textbf{0.95} & \cellcolor{green!25}\textbf{0.97}\\
			
			\midrule
			&&\multicolumn{10}{c}{\textbf{Fully-Supervised Learning}}\\
			\midrule
			\multicolumn{1}{l}{\multirow{3}{*}{XNet\cite{chollet2017xception}}}
			
			&\multicolumn{1}{l}{Precision} & - & - & - & - & - & - & - & - & 0.94 & 0.93\\
			&\multicolumn{1}{l}{Recall}    & - & - & - & - & - & - & - & - & 0.94 & 0.89\\
			&\multicolumn{1}{l}{F1-Score}  & - & - & - & - & - & - & - & - & 0.94 & 0.90\\
			
			\multicolumn{1}{l}{\multirow{3}{*}{InceptionV4\cite{szegedy2016inceptionv4}}}
			&\multicolumn{1}{l}{Precision} & - & - & - & - & - & - & - & - & 0.94 & 0.93\\
			&\multicolumn{1}{l}{Recall}    & - & - & - & - & - & - & - & - & 0.94 & 0.89\\
			&\multicolumn{1}{l}{F1-Score}  & - & - & - & - & - & - & - & - & 0.94 & 0.90\\
			
			\midrule
			&&\multicolumn{10}{c}{\textbf{Semi-Supervised Learning}}\\
			\midrule
			%			\multicolumn{1}{l}{\multirow{3}{*}{ViT-S\cite{vit} LP}}
			%			
			%			&\multicolumn{1}{l}{Precision} & 0.58 & 0.71 & 0.75 & 0.86 & 0.83 & 0.90 & 0.85  & 0.89  & 0.86  & 0.91\\
			%			&\multicolumn{1}{l}{Recall}    & 0.56 & 0.55 & 0.74 & 0.69 & 0.82 & 0.80 & 0.84  & 0.83  & 0.86  & 0.86\\
			%			&\multicolumn{1}{l}{F1-Score}  & 0.55 & 0.60 & 0.74 & 0.74 & 0.82 & 0.83 & 0.84  & 0.85  & 0.86  & 0.87\\
			
			\multicolumn{1}{l}{\multirow{3}{*}{MF-TS\cite{qi2021multi} FT}}
			&\multicolumn{1}{l}{Precision} & 0.77 & 0.86 & 0.83 & 0.92 & 0.91 & 0.96 & 0.93 & 0.97 & - & -\\
			&\multicolumn{1}{l}{Recall}    & 0.77 & 0.81 & 0.82 & 0.87 & 0.91 & 0.96 & 0.93 & 0.96 & - & -\\
			&\multicolumn{1}{l}{F1-Score}  & 0.77 & 0.83 & 0.82 & 0.88 & 0.91 & 0.96 & 0.93 & 0.96 & - & -\\
			
			\bottomrule
		\end{tabular}
	\end{center}
\end{table}

\subsection{Comparison between MoCo-COVID pretrained ViT and ImageNet pretrained ViT}
We evaluated whether MoCo-COVID pretrained ViT-S using original CXR images (CXR-ViT-S) have a better performance compared with model via transfer learning from ImageNet (ImageNet-ViT-S). In the linear probing (LP) protocol, CXR-ViT-S LP (pretrained with MoCo-COVID, only fine-tuning classification layer) with all label fractions achieved statistical significance compared with its ImageNet pretrained counterparts (ImageNet-ViT-S LP), except using 100\% label fraction in the \textit{Test-2} ($ p $<0.05 with paired t-test). In the end-to-end fine-tuning protocol (FT), we found CXR-ViT-S FT (MoCo-COVID pretrained end-to-end ViT-S) outperforms its ImageNet pretrained counterparts (ImageNet-ViT-S FT) more at small label fractions than at large label fractions. With label fraction less than 10\%, CXR-ViT-S FT is statistically better than ImageNet-ViT-S FT ($ p $<0.05 with paired t-test). These observation supports the hypothesis that representations learned in same domain by self-supervised learning are superior than those learned from natural images, and are more significant with scarce of labeled data.

\subsection{Confusion Matrix for MF-ViT CA}
\begin{figure}
	\begin{center}
		\includegraphics[width=\textwidth]{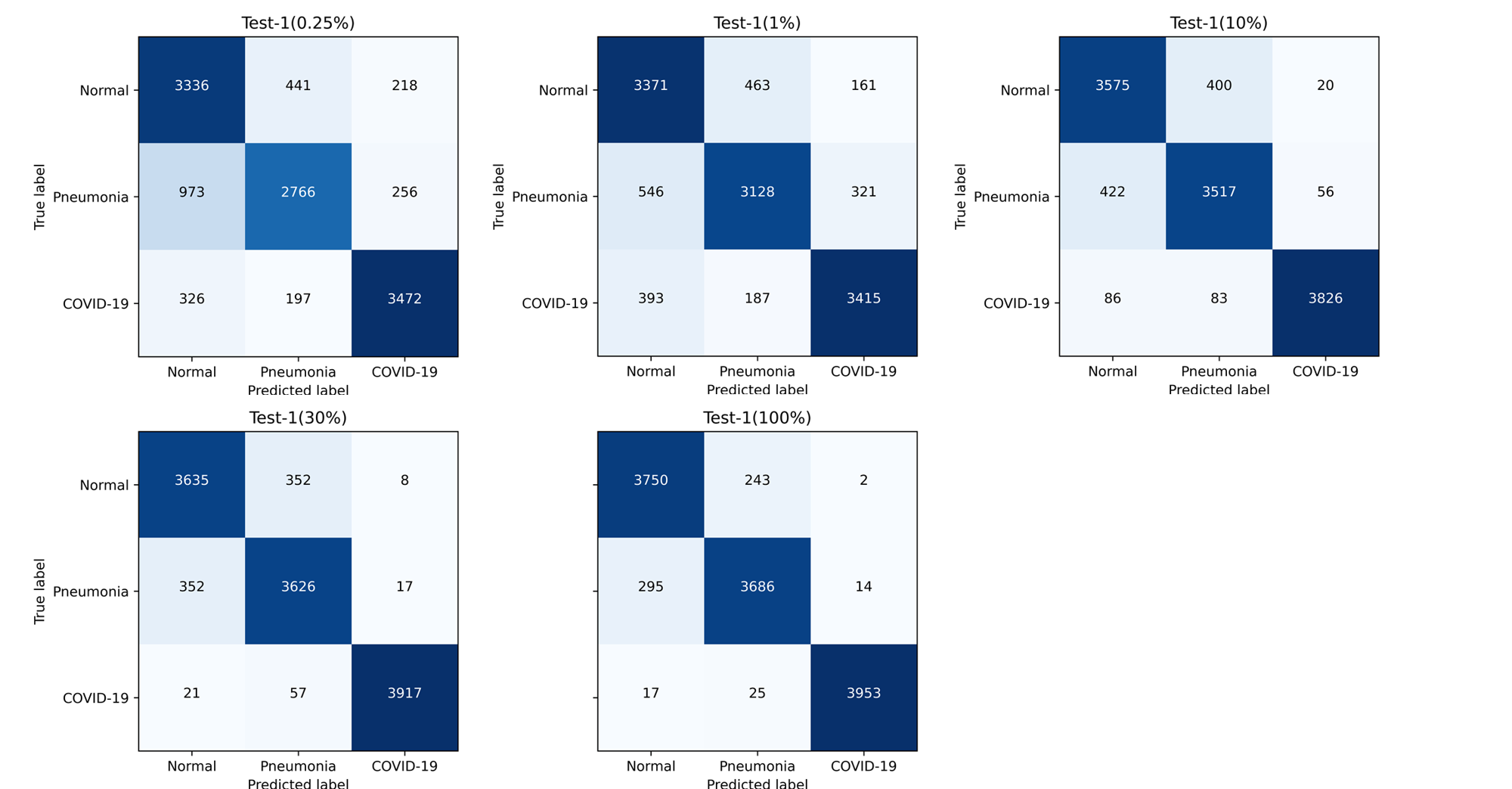}
		\caption{Confusion matrix of MF-ViT CA tested on the \textit{Test-1}.}
		\label{ca_mtr_test1}
	\end{center}
\end{figure}

\begin{figure}
	\begin{center}
		\includegraphics[width=\textwidth]{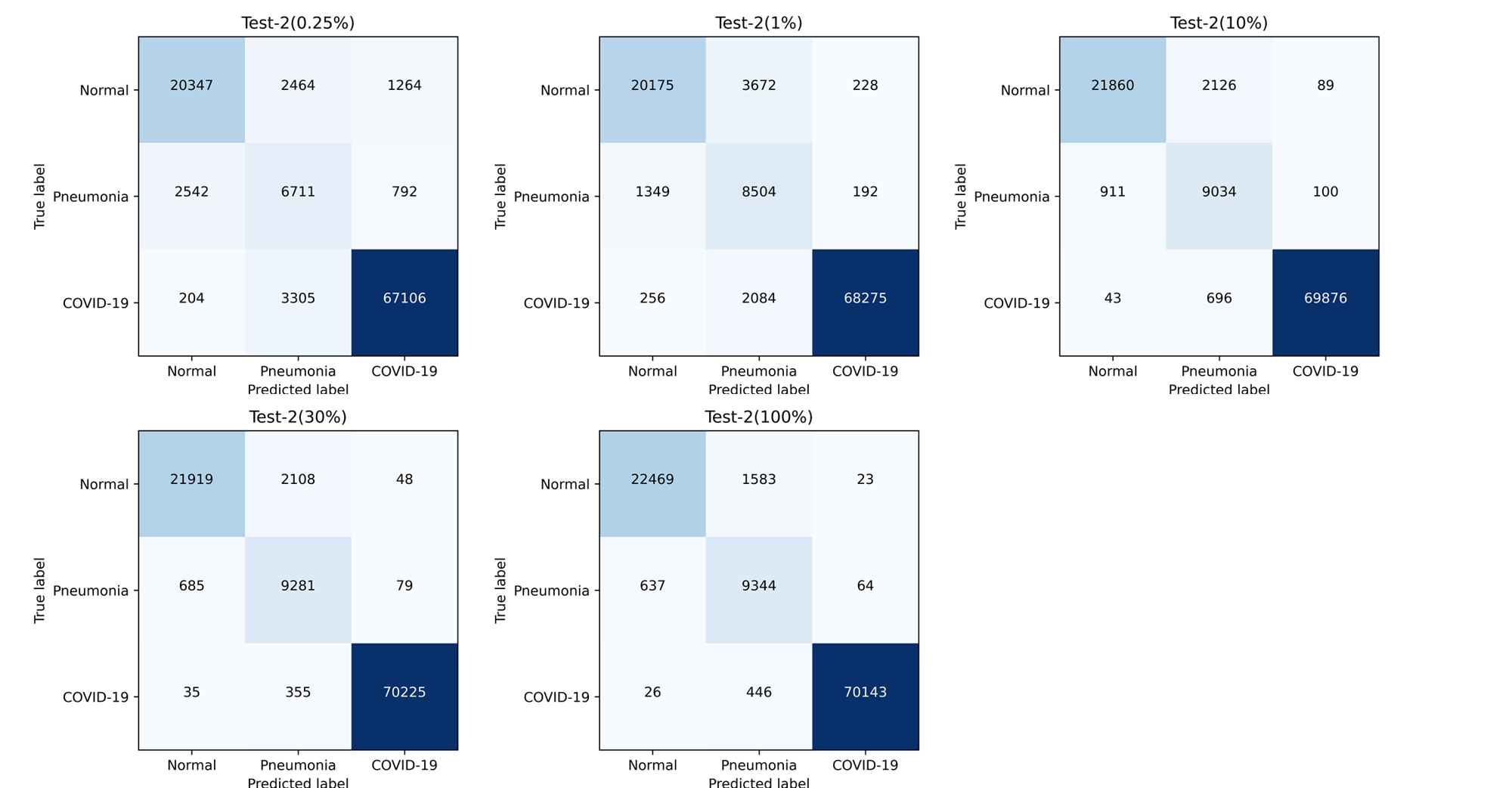}
		\caption{Confusion matrix of MF-ViT CA tested on the \textit{Test-2}.}
		\label{ca_mtr_test2}
	\end{center}
\end{figure}

Figure.\ref{ca_mtr_test1} and Figure.\ref{ca_mtr_test2} shows the confusion matrix of Multi-feature cross-attention model (MF-ViT CA) tested on the \textit{Test-1} and \textit{Test-2} datasets respectively. We found the COVID-19 class has the best performance compared with normal and pneumonia classes across all label fraction.  

\newpage
\bibliographystyle{splncs04}
\bibliography{mybibliography}

\end{document}